\documentclass[12pt,preprint]{aastex}
\newcommand{\Msun}{\ensuremath{\mbox{M}_\odot}}

\newcommand{\gcc}{\ensuremath{\mathrm{g}~\mathrm{cm}^{-3}}}

\newcommand{\cc}{\ensuremath{\mathrm{cm}^{-3}}}
\newcommand{\Hmol}{\ensuremath{\mathrm{H}_2}}

\newcommand{\rarrow}{\ensuremath{\rightarrow}}
\newcommand{\cHI}{\ensuremath{\mathrm{H}}}
\newcommand{\cHII}{\ensuremath{\mathrm{H}^{+}}}
\newcommand{\cHM}{\ensuremath{\mathrm{H}^{-}}}

\newcommand{\celec}{\ensuremath{\mathrm{e}^{-}}}
\newcommand{\cHmolI}{\ensuremath{\mathrm{H}_{2}}}

\newcommand{\cHDI}{\ensuremath{\mathrm{HD}}}

\usepackage{natbib}
\usepackage{graphicx}
\defcitealias{ABN02}{ABN02}
\defcitealias{PSS83}{PSS83}
\defcitealias{2007MNRAS.377..705F}{FH07}
\begin{document}
\title{Effects of Varying the Three-Body Molecular Hydrogen Formation Rate in
Primordial Star Formation}
\author{Matthew J.~Turk\altaffilmark{1}, Paul Clark\altaffilmark{2},
S.~C.~O.~Glover\altaffilmark{2}, T.~H.~Greif\altaffilmark{3}, Tom
Abel\altaffilmark{4}, Ralf Klessen\altaffilmark{2,4}, Volker
Bromm\altaffilmark{5}}
\altaffiltext{1}{Center for Astrophysics and Space Sciences, University of California-San Diego, 9500 Gilman Drive, La Jolla, CA 92093}
\altaffiltext{2}{Zentrum f\"ur Astronomie der Universit\"at Heidelberg, Institut f\"ur Theoretische
Astrophysik, Albert-Ueberle-Str.\ 2, 69120 Heidelberg, Germany}
\altaffiltext{3}{Max-Planck-Institut f\"ur Astrophysik, Karl-Schwarzschild-Strasse 1, 85740 Garching bei M\"uchen, Germany}
\altaffiltext{4}{Kavli Institute for Particle Astrophysics and Cosmology, Stanford University, 2575 Sand Hill Road, Menlo Park, CA 94025, USA}
\altaffiltext{5}{Department of Astronomy, University of Texas, Austin, TX 78712, USA}
\begin{abstract}
The transformation of atomic hydrogen to molecular hydrogen through three-body
reactions is a crucial stage in the collapse of primordial, metal-free halos,
where the first generation of stars (Population III stars) in the Universe are
formed.  However, in the published literature, the rate coefficient for this
reaction is uncertain by nearly an order of magnitude.  We report on the
results of both adaptive mesh refinement (AMR) and smoothed particle
hydrodynamics (SPH) simulations of the collapse of metal-free halos as a
function of the value of this rate coefficient.  For each simulation method, we
have simulated a single halo three times, using three different values of the
rate coefficient.  We find that while variation between halo realizations may
be greater than that caused by the three-body rate coefficient being used, both
the accretion physics onto Population III protostars as well as the long-term
stability of the disk and any potential fragmentation may depend strongly on
this rate coefficient.
\end{abstract}
\maketitle

\section{Introduction}

To constrain the initial mass of the first stars in the Universe, the so-called
Population III stars, we need to be able to understand and model the
gravitational collapse of the progenitor clouds that give birth to them. The
physics of this collapse is governed in large part by the thermal evolution of
the gas \citep[see e.g.][]{2004ARA&A..42...79B,2005SSRv..117..445G}. In
primordial, metal-free gas, this is regulated by cooling from molecular
hydrogen (\cHmolI), as has been known since \citet{1967Natur.216..976S} first
constructed analytical estimates for the importance of molecular hydrogen
cooling in pre-galactic clouds. Early studies of the formation of molecular
hydrogen in protogalaxies focussed on its formation by  ion-neutral reactions at
low densities
\citep{1967Natur.216..976S,1968ApJ...154..891P,1969PThPh..42..219M}. In this
regime, the dominant formation pathway has been shown to be the
electron-catalyzed pair of reactions
\begin{eqnarray}
\cHI + \celec & \rarrow & \cHM + \gamma \label{h2comp:eq:k7} \\
\cHM + \cHI & \rarrow & \cHmolI  + \celec. \label{h2comp:eq:k8}
\end{eqnarray}
The rate-limiting step in this pair of reactions is generally the formation of
the \cHM~ion, and the rate coefficient for this reaction is known very
accurately \citep{1998A&A...335..403G}. The rate coefficient for the second reaction,
the associative detachment of the \cHM~ion to form \cHmolI, is more uncertain,
and in some circumstances, this uncertainty can significantly affect the time
taken for the primordial gas to undergo gravitational collapse and the minimum
temperature reached during the collapse
\citep{2006ApJ...640..553G,2008MNRAS.388.1627G}. However, recent experimental
work \citep{Kreckel10} has reduced this uncertainty to a level at which
it is unlikely to significantly affect the results of future calculations.

This rate at which \cHmolI~can be formed by reactions~\ref{h2comp:eq:k7} and
\ref{h2comp:eq:k8} depends on the free electron abundance. As primordial gas
collapses, recombination causes it to decline, with the result that further
\cHmolI~formation soon becomes very difficult. By comparing the timescales for
\cHmolI~formation and \cHII~recombination, it is simple to show that the
asymptotic fractional abundance of \cHmolI~should be of order $10^{-3}$
\citep[see e.g.][]{1998PThPh.100...63S}, an estimate that has since been
confirmed in numerous numerical simulations (see e.g.~\citet{ABN02}, hereafter
ABN02, and \citet{2002ApJ...564...23B, 2003ApJ...592..645Y}).  However, this is
not the end of the story. At high densities, \citet{PSS83} (hereafter PSS83)
showed that three-body processes would come to dominate the formation of
\cHmolI. A number of different three-body reactions are possible \citep[see the
discussion in][]{2009MNRAS.393..911G}, but the dominant reaction involves
atomic hydrogen as the third body:
\begin{equation}
\cHI + \cHI + \cHI \rightarrow \cHmolI + \cHI.\label{h2comp:eq:k22}
\end{equation}
This reaction has a very small rate coefficient, but the rate at which
\cHmolI~is formed by this process increases rapidly with increasing density.
Therefore, at high densities this reaction is able to rapidly convert most of
the hydrogen in the gas from atomic to molecular form.

The published rate coefficients for reaction~\ref{h2comp:eq:k22} were surveyed
by \citet{2008AIPC..990...25G}, who showed that although most of the published
values agreed reasonably well at high temperatures, they disagreed by orders of
magnitude at the lower temperatures relevant for \cHmolI~formation in
primordial gas. Since \cHmolI~is the dominant coolant in primordial gas, it is
reasonable to suppose that a large uncertainty in its formation rate at high
densities may lead to a large uncertainty in the thermal evolution of this
dense gas. Moreover, the situation is further exacerbated by the fact that each
time an \cHmolI~molecule is formed via reaction~\ref{h2comp:eq:k22},
$4.48~\mathrm{eV}$ of energy is released, corresponding to the binding energy
of the molecule, with almost all of this energy subsequently being converted
into heat. There is thus a substantial chemical heating rate associated with
the three-body formation of \cHmolI, and at the densities at which
reaction~\ref{h2comp:eq:k22} is most important, this can become the dominant
source of heat in the gas.

The influence of the uncertainty in the rate coefficient for
reaction~\ref{h2comp:eq:k22} was studied in \citet{2008MNRAS.388.1627G} and
\citet{2009MNRAS.393..911G}, who confirm that it introduces significant
uncertainty into the thermal evolution of the gas at densities $\rho >
10^{-16}~\gcc$. However, both of these studies involved the use of highly
simplified one-zone models for the gas, in which the gas was assumed to
collapse as if in free-fall, with changes in the temperature having no effect
on the dynamical behaviour. 

In this paper, we present the results of a study that uses high-resolution,
high dynamical range hydrodynamical simulations of Population III star
formation to examine the impact that the uncertainty in the rate coefficient
for reaction~\ref{h2comp:eq:k22} has on both the thermal and the dynamical
evolution of the gas, in order to determine whether this uncertainty will be an
important limitation on our ability to make predictions of the Population III
initial mass function. In order to ensure that our results are not unduly
influenced by our choice of numerical method, we perform simulations using two
very different hydrodynamical codes: the Enzo adaptive mesh refinement (AMR)
code, and the Gadget smooth particle hydrodynamics (SPH) code. Although these
codes have been compared in past studies
\citep{2005ApJS..160....1O,2007MNRAS.374..196R}, this is the first time that
they have been directly compared in a study of Population III star formation. 

The structure of our paper is as follows: first we describe the setup and
chemical model for our simulations in Section~\ref{sec:simulations}; in
Section~\ref{sec:results} we describe the results of our calculations; in
Section~\ref{sec:discussion} we discuss these results and their interplay with
the molecular hydrogen three-body rates; and finally we conclude with a summary
of our findings and a suggestion for a three-body rate to standardize on in
Section~\ref{sec:conclusions}.

%%
%% Methods
%%

\section{Simulations}\label{sec:simulations}

\subsection{Three-Body Rates}

\begin{table}
\begin{tabular}{ll}
\multicolumn{2}{c}{\Hmol~Formation Rates ($\mathrm{cm}^6~\mathrm{s}^{-1}$)} \\ \hline\hline
ABN02 & $1.3\times10^{-32} (T/300)^{-0.38}\hspace{0.3in}(T < 300~\mathrm{K}) $\\
ABN02 & $1.3\times10^{-32} (T/300)^{-1.00}\hspace{0.3in}(T > 300~\mathrm{K}) $\\
PSS83 & $5.5\times10^{-29}/ T $\\
FH07 & $1.44\times10^{-26} / T^{1.54} $\\ \\
\multicolumn{2}{c}{\Hmol~Destruction Rates ($\mathrm{cm}^3~\mathrm{s}^{-1}$)} \\ \hline\hline
ABN02  & $
(1.0670825\times10^{-10}\times(T_{\mathrm{eV}})^{2.012})/(\exp(4.463/T_{\mathrm{eV}})\times(1+0.2472 T_\mathrm{eV})^{3.512})
$\\
PSS83  & $5.24\times10^{-7}\times T^{-0.485} \exp(-52000/T) $\\
FH07 & $1.38\times10^{-4}\times T^{-1.025} \exp(-52000/T) $
\end{tabular}
\caption{Formation and destruction rates for all three studied sets of rates.
$T$ is the gas temperature in Kelvin and $T_{\mathrm{eV}}$ is defined as $T /
11605$ (the temperature in units of eV).}\label{h2comp:tab:rates}.
\end{table}

\begin{figure*}
\begin{centering}
\plotone{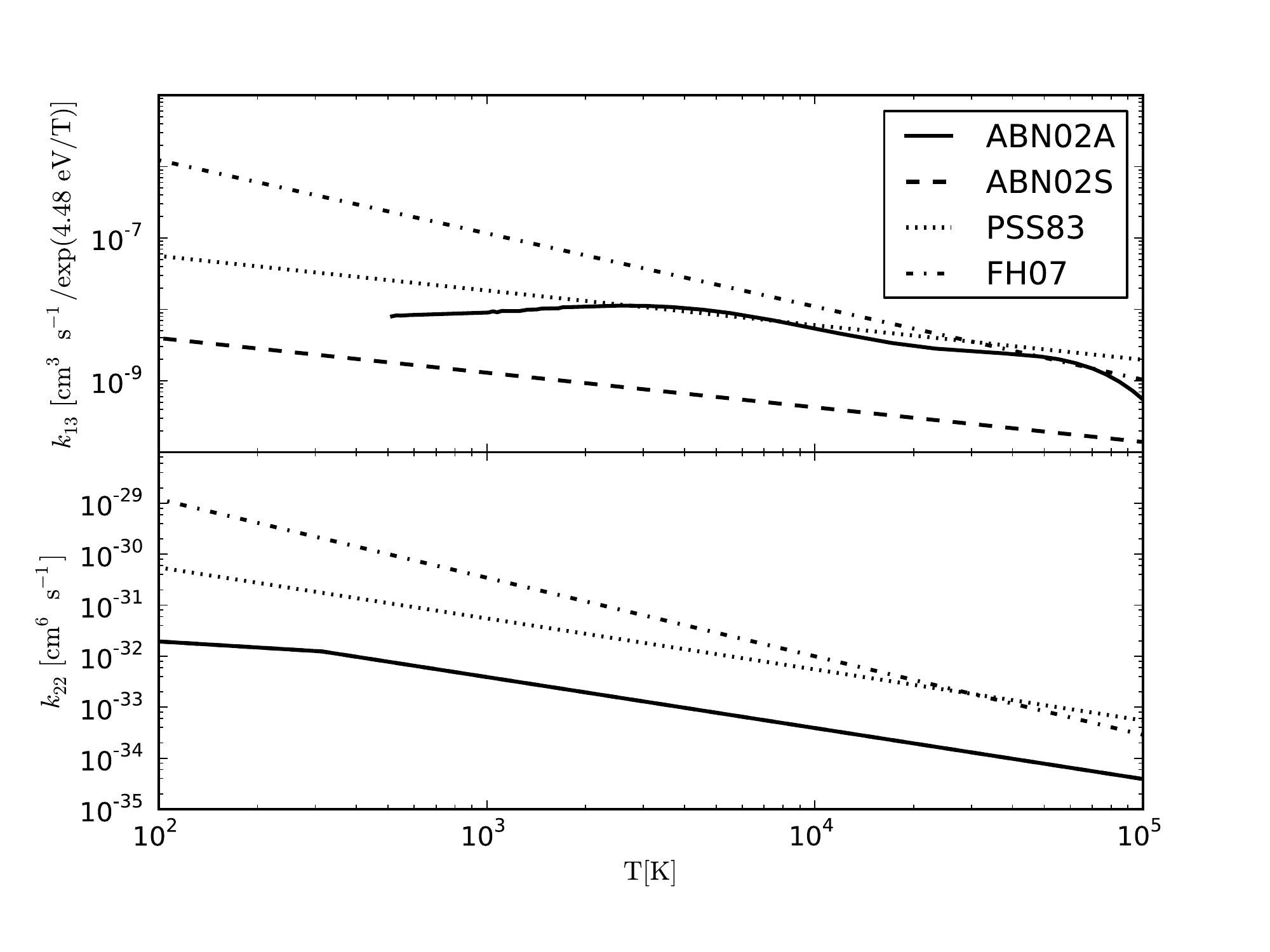}
\caption{Collisional dissociation (top) and three-body association 
(bottom) rates from \citetalias{ABN02} (solid: simulation ABN02A, dashed: ABN02S),
\citetalias{PSS83} (dotted: simulations PSS83A \& PSS83S), and
\citetalias{2007MNRAS.377..705F} (dash-dot: simulations FH07A \& FH07S).  A factor of
\ensuremath{\exp{4.48~\mathrm{eV}/\mathrm{T}}} has been divided out of the
collisional dissociation rate, for clarity.  We note that the solid and dashed
lines, for simulations ABN02A and ABN02S, are coincident in the lower panel.}
\label{h2comp:fig:frate_comp}
\end{centering}
\end{figure*}

The three-body rates we chose to compare in this study were taken from
\citetalias{ABN02}, \citetalias{PSS83}, and
\citet{2007MNRAS.377..705F} (hereafter FH07); their values as a function of
temperature are shown in Figure~\ref{h2comp:fig:frate_comp}.  We have tabulated
these rates in Table~\ref{h2comp:tab:rates}.  The \citetalias{ABN02} rate is
based on the theoretical calculations of \cite{1987JChPh..87..314O} at low
temperatures ($T < 300 \: {\rm K}$). At higher temperatures, \citeauthor{ABN02}
assumed, in the absence of better information, that the rate was inversely
proportional to temperature. This choice means that the \citeauthor{ABN02} rate
has a sudden change of slope at $T = 300 \: {\rm K}$, which is somewhat
artificial. However, in practice this feature appears to be harmless for
Population III.1 star formation, as in previous simulations of collapsing
primordial gas clouds, the gas was always significantly hotter than 300~K by the
time the gas density reached the domain in which three-body \Hmol~formation
dominates.  For subsequent Population III.2 star formation, where \cHDI cooling may
dominate, this feature may indeed be important \citep{2007ApJ...667L.117Y}.  The
AMR and SPH calculations utilized different dissociation rates for the ABN02
calculations, which have been plotted separately; the AMR calculation utilized a
density-dependent match to the rate given in \cite{MSM96}, whereas the SPH
calculation utilized a temperature-dependent value calculated via the principle
of detailed balanced.  We have plotted the ABN02A rate at a density of
$\mathrm{n}_{\mathrm{H}} = 10^9~\mathrm{cm}^{-3}$.  The rates from
\citetalias{PSS83} and \citetalias{2007MNRAS.377..705F} were both computed using
the principle of detailed balance, applied to the two-body collisional
dissociation reaction

\begin{equation}
\cHmolI + \cHI \rarrow  \cHI + \cHI + \cHI.  \label{inverse}
\end{equation}

The two studies obtained different three-body rates from this procedure on
account of the different assumptions they made regarding the temperature
dependence of the \Hmol~partition function \citep{2007MNRAS.377..705F}. In all
three cases, we ensured that the \Hmol~collisional dissociation rate used in
the simulations was consistent with the chosen three-body \Hmol~formation rate.
We know that for a system in chemical and thermal equilibrium, the rate at which 
H$_{2}$ is produced by reaction~\ref{h2comp:eq:k22} must equal the rate at which 
it is destroyed by reaction~\ref{inverse}; 
this is a simple consequence of the principle of microscopic reversibility (see 
e.g. \citealt{Denbigh} for a detailed discussion). We also 
know that for a system in chemical and thermal equilibrium, the ratio of the
equilibrium abundances of atomic and molecular hydrogen are related by the Saha
equation
\begin{equation}
\frac{n_{\rm H_{2}}}{n_{\rm H}^{2}} = \frac{z_{\rm H_{2}}}{z_{\rm H}^{2}} 
\left(\frac{h^{2}}{\pi m_{\rm H} k T} \right)^{3/2} \exp \left(\frac{E_{\rm diss}}{kT}
\right),
\end{equation}
where $n_{\rm H_{2}}$ and $n_{\rm H}$ are the number densities of H$_{2}$ and atomic
hydrogen, respectively, $z_{\rm H_{2}}$ and $z_{\rm H}$ are the partition functions
of H$_{2}$ and atomic hydrogen, $E_{\rm diss}$ is the dissociation energy of H$_{2}$
and the other symbols have their usual meanings. Now, since $k_{\rm f} / k_{\rm r}
= n_{\rm H_{2}} / n_{\rm H}^{2}$, where $k_{f}$ is the rate coefficient for 
reaction~\ref{h2comp:eq:k22} and $k_{r}$ is the rate coefficient for 
reaction~\ref{inverse}, this implies that
\begin{equation}
\frac{k_{\rm f}}{k_{\rm r}} = \frac{z_{\rm H_{2}}}{z_{\rm H}^{2}} 
\left(\frac{h^{2}}{\pi m_{\rm H} k T} \right)^{3/2} \exp \left(\frac{E_{\rm diss}}{kT}
\right),
\label{eq_constant}
\end{equation}
provided that the gas in is local thermodynamic equilibrium (LTE), which is a reasonable
approximation at the densities and temperatures for which three-body H$_{2}$ formation
is an important process. Therefore, if we change the three-body H$_{2}$ formation rate
coefficient, we must also change the rate coefficient for \Hmol~collisional dissociation
(in the LTE limit) in such a fashion that Equation~\ref{eq_constant} remains satisfied. 

Two additional rates discussed in \cite{2008AIPC..990...25G} -- one that was
derived there for the first time, and another that was suggested by \cite{CW83}
-- are not included in our study.  In common with the \citetalias{PSS83} rate, these
rates lie between the extremes represented by the \citetalias{ABN02} and
\citetalias{2007MNRAS.377..705F} rates, and so we would expect them to yield
behaviour similar to that we find for the \citetalias{PSS83} rate.

Finally, we note that in this study we do not investigate the effects of any uncertainties in
the rate coefficients of other three-body reactions, such as
\begin{equation}
{\rm H} + {\rm H} + {\rm He} \rightarrow {\rm H_{2}} + {\rm He}.
\end{equation}
This process is included in the chemical network used for our SPH simulations, along with
its inverse
\begin{equation}
{\rm H_{2}} + {\rm He} \rightarrow {\rm H} + {\rm H} + {\rm He}, 
\end{equation}
but in practice their effects are unimportant, as these reactions are never responsible for more than 
a few percent of the total H$_{2}$ formation or destruction rate (see e.g.\ \citealt{2009MNRAS.393..911G}). 
These reactions are not included in the chemical network used for our AMR simulations, but in view
of their unimportance, we do not expect this difference in the chemical networks to significantly affect 
our results.

\begin{figure}
\begin{centering}
\plottwo{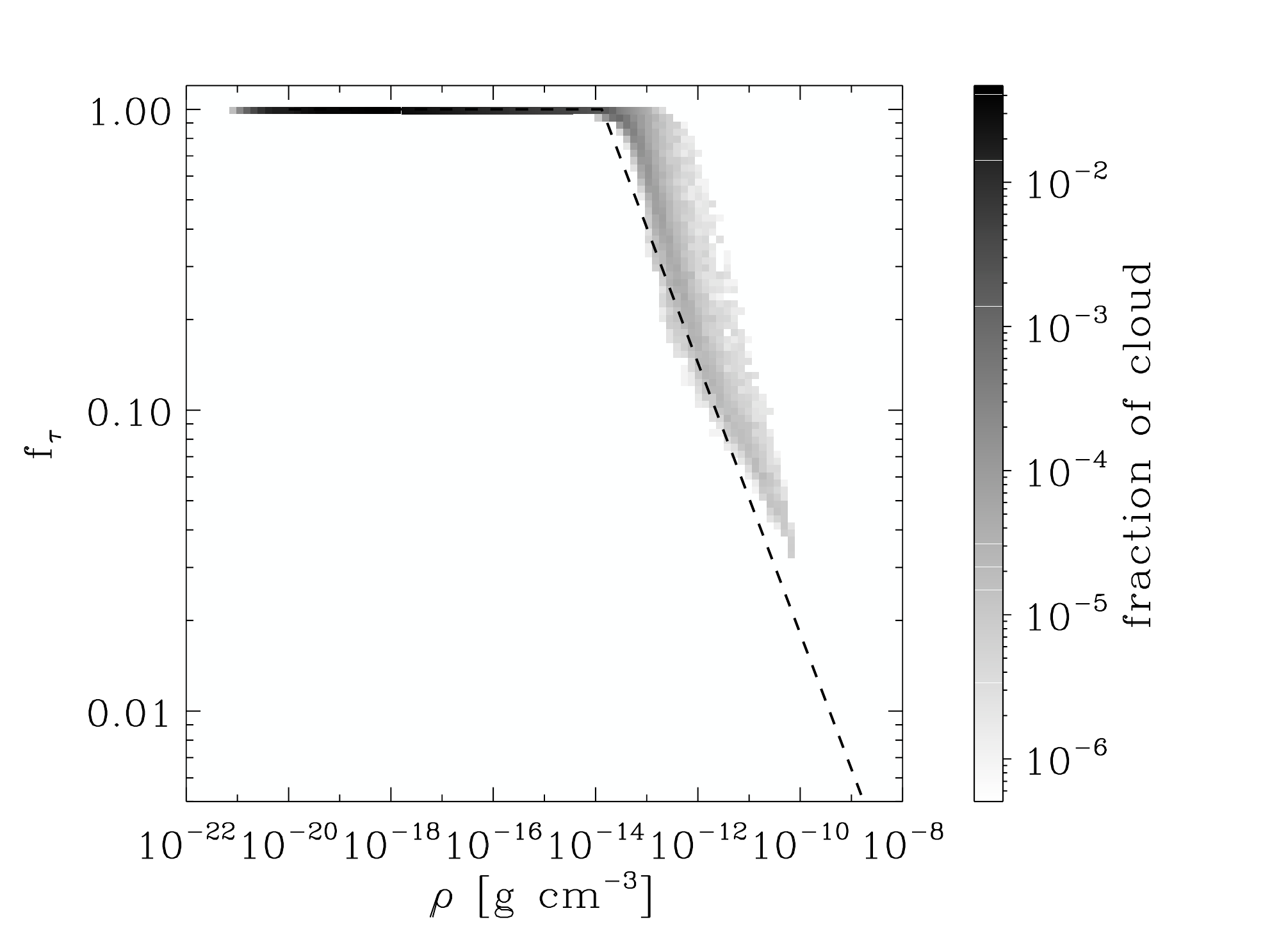}{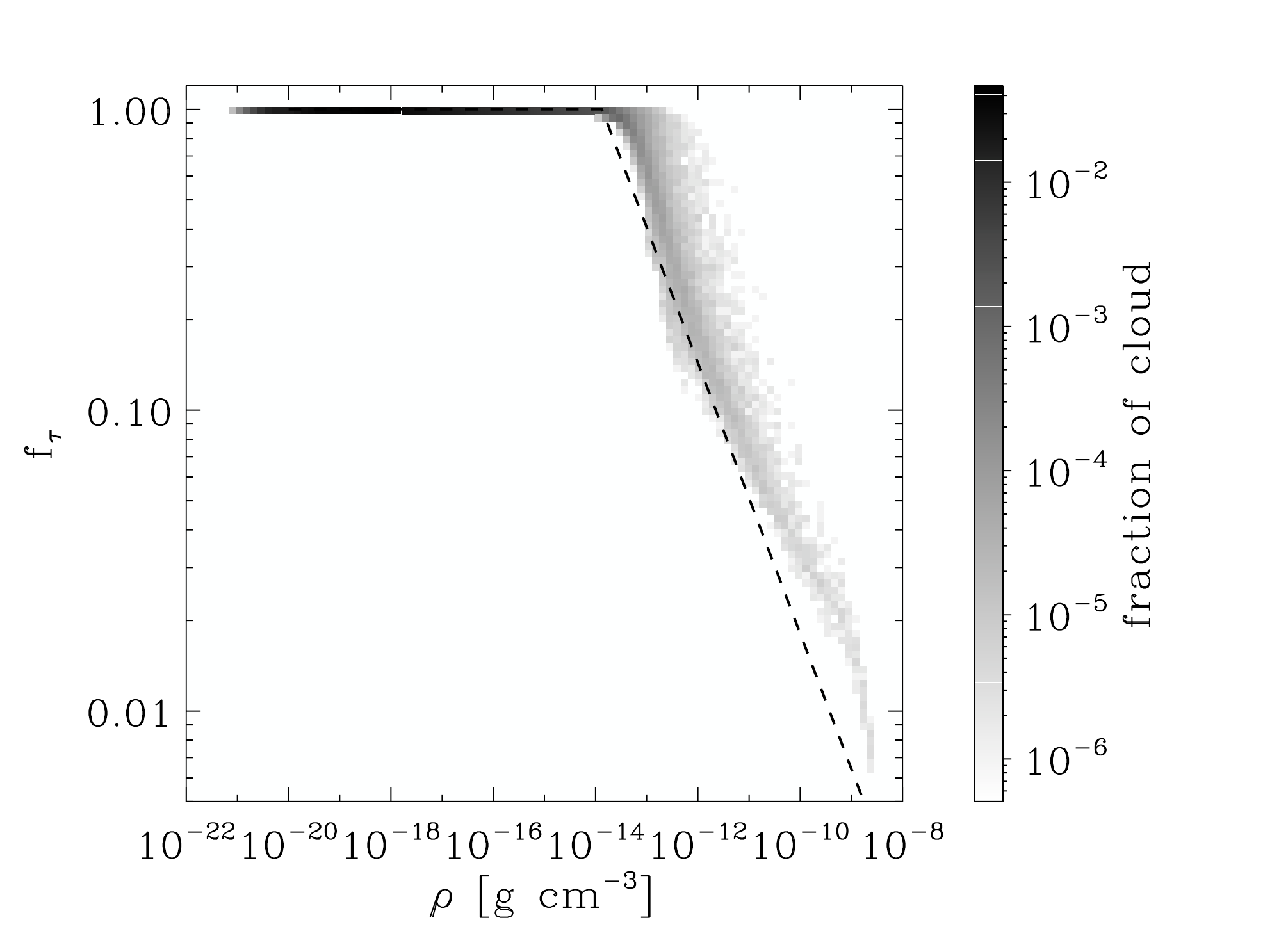}
\caption{Comparison of the optical depth approximations for the SPH
calculations (phase diagram) versus the AMR calculations (solid black line).
$f_{\tau}$ is the ratio between the optically thick and optically thin cooling rates.
The left-hand panel compares the two approximations at the time at which the
peak density was $7.9 \times 10^{-11}\gcc$, while the right-hand panel
shows the comparison at a later time, when the peak density was
$2.7 \times 10^{-9} \gcc$.}
\label{h2comp:fig:ftau_comp}
\end{centering}
\end{figure}

\subsection{Simulation Setup}
For each of our selected three-body rates, we perform two
simulations, one using an adaptive mesh refinement (AMR) code, and a second one
using a smoothed particle hydrodynamics (SPH) code. As well as utilizing
different computational approaches, these two sets of simulations also use
slightly different initial conditions, as described below, although in both
cases the simulations probe what are typical conditions for the formation of
Population III stars. Our rationale for this dual approach is to be able to
demonstrate that the uncertainty introduced into the outcome of the simulations
by our poor state of knowledge regarding the rate of
reaction~\ref{h2comp:eq:k22} is largely independent of our choice of initial
conditions or numerical method.  In the following subsections, we describe in
more detail the setups used for our AMR and SPH simulations
(sections~\ref{amr-method} and \ref{sph-method}, respectively), and also
briefly discuss the different approaches that we use to treat \Hmol~cooling in
the optically thick limit (section~\ref{op-thick}).

\subsubsection{Adaptive mesh refinement simulations}
\label{amr-method}
Simulations ABN02A, PSS83A and FH07A were conducted using the adaptive mesh
refinement code Enzo, a three-dimensional cosmological adaptive mesh refinement
code written by Greg Bryan, with ongoing development at many institutions,
including the Laboratory for Computational Astrophysics (UCSD), Stanford
University, and Columbia University.  Enzo includes physical models for
radiative cooling, non-equilibrium primordial chemistry, $N$-body dark matter
and hydrodynamics (\citealt{oshea04}; \citealt{2001astro.ph.12089B};
\citealt{1997astro.ph.10187B}, data analysis using \texttt{yt},
\citealt{yt_full_paper}).  Here, we use a coupled chemistry and cooling
solver, including the $4.48\mathrm{eV}$ energy deposition and removal for each
molecule of hydrogen.  Each simulation uses a single set of rate coefficients,
as noted in Figure \ref{h2comp:fig:frate_comp}.  ABN02A uses the rates taken
from \citetalias{ABN02}, PSS83A from \citetalias{PSS83}, and FH07A from
\citetalias{2007MNRAS.377..705F}.  

Simulations ABN02A, PSS83A, and FH07A were initialized at $z=99$ assuming a
concordance $\Lambda$ cold dark matter ($\Lambda$CDM) cosmological model:  
total matter density $\Omega_{\mbox{\tiny{m}}} = 0.27$, baryon density 
$\Omega_{\mbox{\tiny{b}}} = 0.0463$, dark matter density 
$\Omega_{\mbox{\tiny{CDM}}} = 0.2237$, dark energy density 
$\Omega_\Lambda = 0.73$, Hubble parameter 
$h=H_{0}/100~{\rm km}~{\rm s}^{-1}~{\rm Mpc}^{-1}=0.72$, where $H_{0}$
is the Hubble expansion rate today,  spectral index $n_{s}=1.0$,
and power spectrum normalization $\sigma_8 = 0.7$ \citep{Spergel06}.
However, while the specific cosmological parameters govern the halo
itself, we intend to compare the hydrodynamical and chemical state of the gas
across realizations while varying the formation and dissociation rate of
molecular hydrogen; the results should be largely immune to small variations in
the cosmology used.  A random cosmological realization is used, with a box size
of $0.3h^{-1}$~Mpc (comoving), centered on the location of the earliest
collapsing massive halo, of mass $5\times10^5~\Msun$.  We use recursive
refinement to generate higher-resolution subgrids, with an effective resolution
of $1024^3$ in the region of collapse.  The most massive halo collapses at
$z=17.1$, and we halt each of the four simulations when the maximum number
density is $10^{16}~\cc$, corresponding to a mass density slightly greater than
$10^{-8}~\gcc$.

\subsubsection{Smoothed particle hydrodynamics simulations}
\label{sph-method}
Simulations ABN02S, PSS83S and FH07S were conducted using the Gadget 2
smoothed particle hydrodynamics code \citep{2005MNRAS.364.1105S}. 
We have modified the publicly available version of Gadget 2 to add a treatment 
of primordial gas chemistry and cooling, discussed in detail elsewhere 
\citep{2007ApJ...666....1G,Clark2010S}.
As with the AMR simulations, each of our SPH simulations uses a single set
of rate coefficients, taken from \citetalias{ABN02}, \citetalias{PSS83} and 
\citet{2007MNRAS.377..705F}, respectively.

The SPH simulations differ from the AMR in that they are run in two distinct
stages. First, we model the formation of the
minihalo in a cosmological simulation. We choose a side length of $200~{\rm
kpc}$ (comoving) and initialize the parent simulation at $z=99$ with a
fluctuation power spectrum determined by a concordance $\Lambda$CDM 
cosmology with $\Omega_{m}=0.3$, $\Omega_{b}=0.04$, 
$h=0.7$, $n_{s}=1.0$, and $\sigma_{8}=0.9$ \citep{Spergel2003}.
In a preliminary run with $128^3$ dark matter particles and $128^3$ gas particles, 
we locate the formation site of the first minihalo that collapses and cools to high densities. 
We then re-initialize the simulation with three consecutive levels of refinement,
replacing a parent particle by a total of $512$ daughter particles. To avoid
the propagation of numerical artifacts caused by the interaction of particles
of different masses, we choose the highest resolution region to have a side
length of $50~{\rm kpc}$, which is much larger than the comoving volume of the
minihalo. The particle mass within this region is $\simeq 0.26~\rm{M}_{\odot}$
in DM and $\simeq 0.04~{\rm M}_{\odot}$ in gas.

The cosmological simulation is then evolved until the gas in the minihalo has
reached a density of $\rho = 10^{-18} \: \mathrm{g} \: \mathrm{cm}^{-3}$, by which point the 
gas has gravitationally
decoupled from its parent minihalo and has begun to collapse in its own right.
At this point we discard the full cosmological simulation and focus our
calculation on the central collapsing region and its immediate surroundings.
These then become the initial conditions for our study of the three-body formation
rates. The initial gravitational instability that leads to the collapse in the
baryons occurs at around $\rho \sim 10^{-20}~\gcc$ and $T \sim 270$ K, which
corresponds to a Jeans mass of around 350 $\Msun$. To ensure that we capture
the entire collapsing fragment in our simulations, and to avoid any unphysical
boundary effects, we select a spherical region containing 1000~$\Msun$. To
account for the effects of the missing gas  that should surround our central
core, we include a external pressure \citep{1990nmns.work..269B} that modifies
the standard gas-pressure contribution to the Gadget2 momentum equation, 

\begin{equation}
\frac{d v_{i}}{d t} = - \sum_{j} m_{j} \left[
f_{i}\frac{P_{i}}{\rho_{i}^{2}} \nabla_{i} W_{ij}(h_{i})
+ f_{j}\frac{P_{j}}{\rho_{j}^{2}}\nabla_{i} W_{ij}(h_{j}) \right] , 
\end{equation}

\noindent by replacing $P_{i}$ and $P_{j}$ with $P_{i} -  P_{\rm ext}$ and
$P_{j} -  P_{\rm ext}$  respectively, where $P_{\rm ext}$ is the external
pressure, and all quantities have the usual meaning, consistent with those used
by \citet{2005MNRAS.364.1105S}. The pair-wise nature of the force summation
over the SPH neighbors ensures that $P_{\rm ext}$ cancels for particles that
are surrounded by other particles. Only at the edge does the term not
disappear, where it mimics the pressure contribution from the surrounding
medium. The average density and temperature at the edge of our $1000~\Msun$
cloud are $10^{-20}~\gcc$ and 270 K respectively.  These average values are
used to define the value of $P_{\rm ext}$.

To evolve the collapse of the gas to high densities, we also need to increase
the resolution. Since the SPH particle mass in the original cosmological
simulation was $0.04~\Msun$, our selected region contains only $\sim 20,000$
SPH particles. To increase the resolution, we `split' the particles into 100
SPH particles of lower mass. This is done by randomly placing the sibling
particles inside the smoothing length of parent particle. Apart from the mass
of the siblings, which is 100 times less that of the parent, they inherit the
same values for the entropy, velocities and chemical abundances as their
parents.

Although this set-up permits us to follow the evolution of the baryons over
many orders of magnitude in density, the SPH calculations in this study do not
achieve the same resolution as the AMR calculations. The initial Jeans-unstable
region in the SPH calculations is resolved by roughly 9,000 particles, or $\sim
20^{3}$. In contrast, the AMR calculations are set to resolve the Jeans length
by $16^{3}$ grid cells, {\em at all times}.  As such, they can much better
resolve the turbulence and structure that develops during the collapse of the
baryons. Along with the slightly higher degree of rotation found in the minihalo
modelled in the SPH calculations, this explains why the SPH calculation contain
significantly less density structure than the AMR simulations. This will be
discussed further in Section~\ref{sec:results}.

\subsubsection{Molecular hydrogen cooling in optically thick gas}
\label{op-thick}
The two different codes utilized different mechanisms for treating
\Hmol~cooling at high densities  ($\rho \geq 10^{-14}\gcc$), in the regime
where the lines become optically thick.  In the AMR simulations, we used the
approximation proposed by \cite{RA04}, which provides a functional form
dependent only on density, initially calculated in that work via escape
fraction estimates and then calibrated to 1D simulation results.  In the SPH
simulations, we used the classic Sobolev approximation, as implemented in
\cite{2006ApJ...652....6Y}, where the local density and velocity gradient are
used to compute an optical depth correction independently for each gas
particle. In Figure~\ref{h2comp:fig:ftau_comp}, we compare the effective 
suppression of ro-vibrational cooling produced by these two approaches 
at two different times during the collapse, using data from run ABN02S to
construct the optical depth correction factors for the Sobolev approach.
We note that while the two approximations are in general agreement, the 
\cite{RA04} approximation generally suppresses cooling slightly more than the 
Sobolev-based approximation, albeit with a sharper turn-on point. Comparison
of the left and right-hand panels in Figure~\ref{h2comp:fig:ftau_comp} 
demonstrates that the differences between the two methods do not appear
to depend strongly on the time at which the comparison is made, although it
is plausible that we would find greater differences were we to examine much 
later times in the evolution of the system, after the formation of the initial protostar.
The differences between the two methods for suppressing ro-vibrational cooling may 
affect the temperature of the two simulations, but as they are typically not 
larger than a factor of two,  we believe that this effect will be secondary to the variance 
in the physical conditions between the two halos. 

\section{Results}\label{sec:results}

Except where otherwise indicated, we have compared the two sets of simulations at
the time when their peak density was $10^{-8}~\gcc$.  Ideally, these simulations
would be compared at epochs of relative time between collapses and transitions
between physical processes; however, owing to the difference in halos, the
difference in collapse time (\S~\ref{sec:timing}) and pragmatic issues of
coordinating two different simulation methodologies and two different halos, we
have instead chosen to compare the six simulations at identical peak densities.

\subsection{Collapse Time}
\label{sec:timing}
We first investigated the time taken to reach a density of  $10^{-8}~\gcc$ in
the six simulations. To allow us to directly compare the final SPH simulations,
which start with a central density of $3.4 \times 10^{-18}\gcc$, with the AMR
simulations, which start from cosmological densities, we chose to measure the
collapse times from the moment at which the maximum density in the simulation
was $10^{-17}~\gcc$. Once we had identified the most rapidly collapsing AMR and
SPH simulations, we computed the time by which collapse was delayed in the more
slowly collapsing simulations. These values are listed in
Table~\ref{h2comp:table:tdeltat}. In both the SPH and the AMR simulations, the
rate of collapse is directly related to the rate coefficient of the 3-body
\Hmol~formation, with the simulations employing the rate coefficient of
\citet{2007MNRAS.377..705F} collapsing the fastest and those using the
coefficient from ABN02 collapsing the slowest.  For reference, the
gravitational free-fall time of the gas with $\rho = 10^{-17}~\gcc$ was
$t_{\mathrm{ff}} = \sqrt{3 \pi / 32 G \rho} \simeq 21000 \: \mathrm{yr}$.  In
addition, the free-fall times at densities of $10^{-16}~\gcc$ and
$10^{-10}~\gcc$ -- those over which the \Hmol~formation takes place -- are 6657
yr and 6.7 yr respectively. As such, the delays seen in the different
simulations are long compared to the free fall times in the relevant stages of
the collapse.  This is particularly apparent in run ABN02A, where the collapse
is delayed by almost seven free-fall times compared to run FH07A.  Any delay in
collapse time can have a profound implication on the incidence of fragmentation
and the formation of multiple stars, as discussed in
Section~\ref{sec:discussion}.

\begin{table}
\begin{center}
\begin{tabular}{lr|lr}
\tableline
Simulation & $\Delta t \: [\mathrm{years}]$ & Simulation & $\Delta t \: [\mathrm{years}]$  \\ \tableline
ABN02A & $ 135447 $ & ABN02S & $ 17908 $ \\
PSS83A & $   6424 $ & PSS83S & $ 15978 $ \\
FH07A &  $      0 $ & FH07S &  $    0 $ \\ \tableline
\end{tabular}
\end{center}
\caption{Time required to collapse from $\rho_{\mathrm{max}}(t) = 10^{-17}~\gcc$
to $\rho_{\mathrm{max}}(t) = 10^{-8}~\gcc$, relative to the the time taken in the most
rapidly collapsing simulation. 
\label{h2comp:table:tdeltat}}
\end{table}

\subsection{Radial Profiles}

\begin{figure*}
\plotone{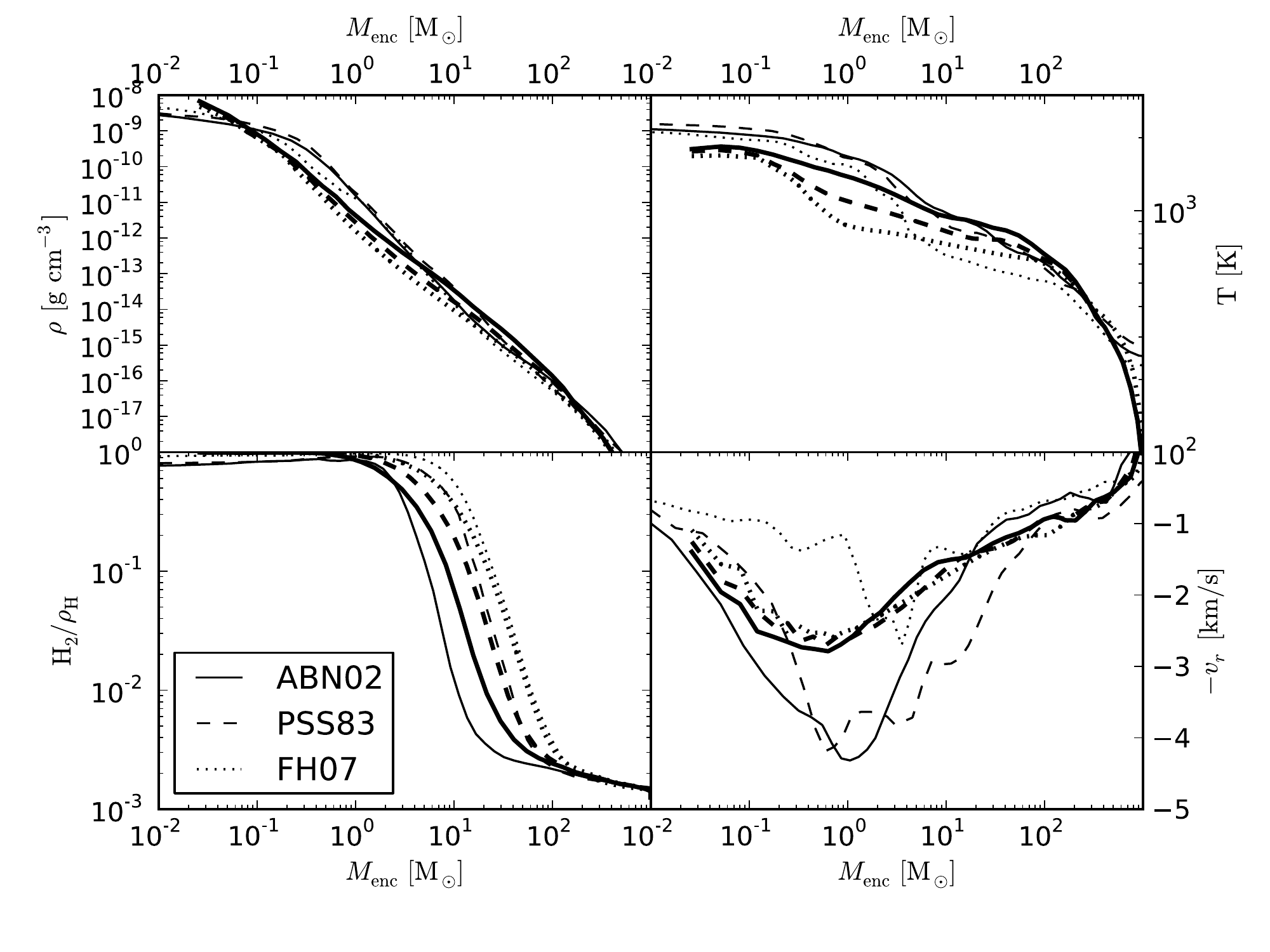}
\caption{Radially-binned, spherically averaged profiles of density
(volume-weighted, upper left), temperature (mass-weighted, upper right),
molecular hydrogen mass-fraction (mass-weighted, lower left) and radial
velocity (mass-weighted, lower right) plotted as a function of enclosed mass,
measured from the densest point in the calculation.  Thin lines correspond to the
AMR simulations and thick lines correspond to the SPH simulations.}
\label{h2comp:fig:menc_multi}
\end{figure*}

\begin{figure*}
\plotone{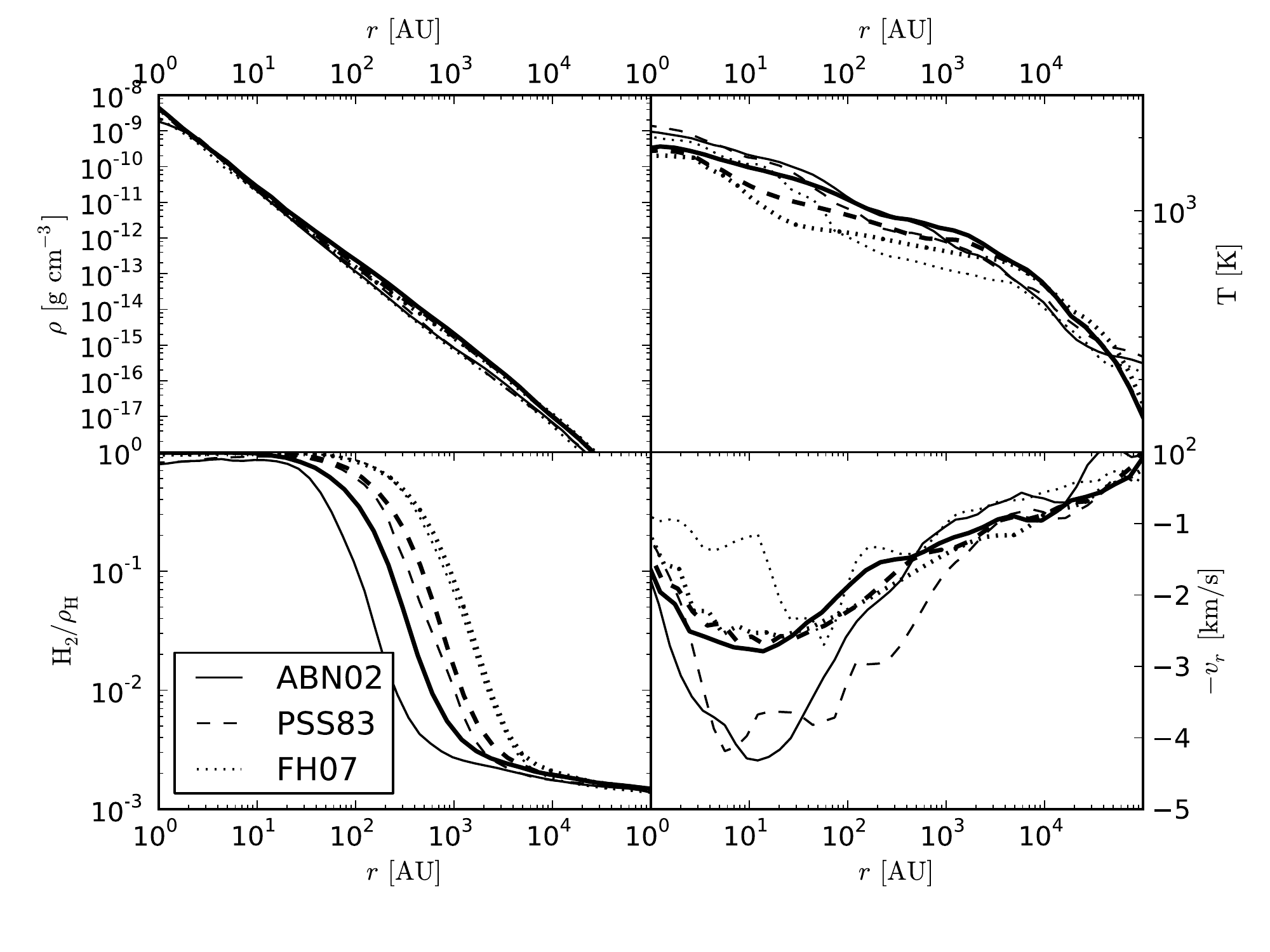}
\caption{Radially-binned, spherically averaged profiles of density
(volume-weighted, upper left), temperature (mass-weighted, upper right),
molecular hydrogen mass-fraction (mass-weighted, lower left) and radial
velocity (mass-weighted, lower right) plotted as a function of radius from the
most dense point in the calculation.  Thin lines correspond to the AMR
simulations and thick lines correspond to the SPH simulations.}
\label{h2comp:fig:radius_multi}
\end{figure*}

\begin{figure*}
\plotone{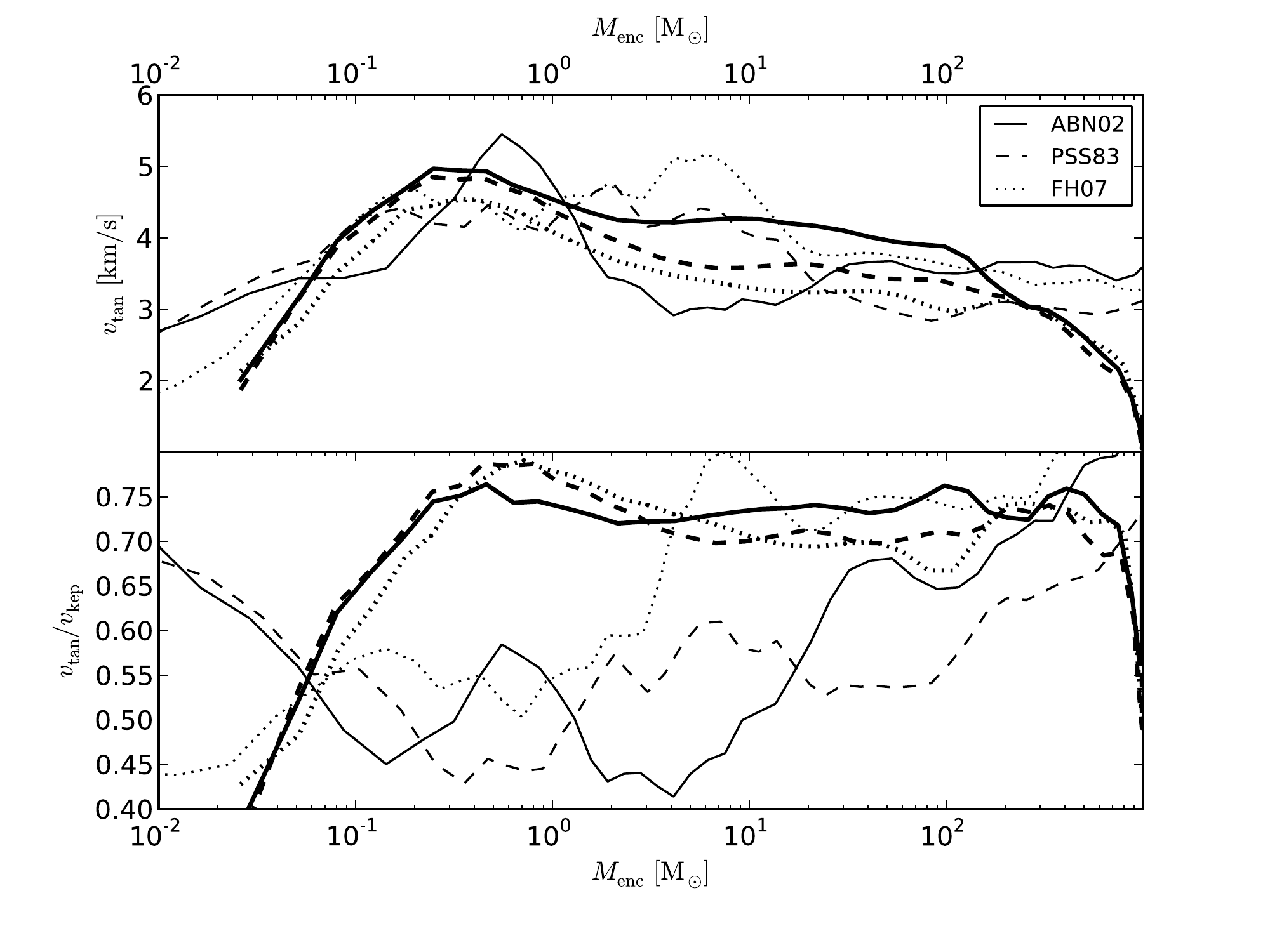}
\caption{Radially-binned, spherically averaged profiles of tangential velocity
(upper panel) and tangential velocity divided by Keplerian velocity (lower
panel), plotted as a function of enclosed mass, measured from the densest 
point in the calculation. The velocity of the innermost bin has been subtracted 
prior to the calculation of the relative velocities.  Thin lines correspond to the AMR
simulations and thick lines correspond to the SPH simulations.}
\label{h2comp:fig:menc_multi_vel}
\end{figure*}

In Figure~\ref{h2comp:fig:menc_multi} we have plotted averaged spherical shell
profiles of the six realizations, where the innermost bin is taken at the most
dense point in the simulation and the values have been plotted as a function of
the mass enclosed within each radial bin.  In
Figure~\ref{h2comp:fig:radius_multi}, we have plotted the same
spherically-averaged values as a function of the radius.

The upper left panels of Figures~\ref{h2comp:fig:menc_multi} and
\ref{h2comp:fig:radius_multi} show the volume-weighted average density, with
the AMR results plotted in thin lines, and the SPH results in thick lines.  In
both sets of simulations, we see a similar dependence on the three-body
reaction rate.  The highest density at a given enclosed mass (or,
alternatively, the highest enclosed mass at a given density) is produced by the
ABN02 and PSS83 rates, which give very similar results. The FH07 rate produces
systematically lower enclosed masses in both simulations, over a wide range in
radii. The differences appear small, owing to the wide range of scales covered
by the plot, but in the worst case can amount to a factor of two uncertainty in
the enclosed mass (see e.g.\ $M_{\mathrm{enc}}$ in the three SPH simulations at
a density $\rho = 10^{-13}~\gcc$). Nevertheless, the uncertainty introduced by
the choice of three-body rate coefficient is less significant than the
difference between the SPH and AMR realizations. For enclosed masses less than
about 10~M$_{\odot}$, the SPH simulations are of characteristically lower
density, indicating a core that is overall less massive. For example, the AMR
simulations enclose roughly $0.5~\Msun$ of gas that is of density
$5\times10^{-10}\gcc$ or higher, whereas the SPH simulations enclose only
$0.1~\Msun$ in this density range.  At enclosed masses greater than
$\sim5~\Msun$, the two sets of simulations are in good agreement, and the
radial profiles show extremely good agreement for all six realizations.

The upper right panels of Figures~\ref{h2comp:fig:menc_multi} and
\ref{h2comp:fig:radius_multi} show the mass-weighted average temperature, where
the AMR simulations are plotted in thin lines and the SPH simulations are
plotted in thick lines.  Here we again see the same trend with rate coefficient
in both sets of simulations: the ABN02 rate produces the hottest gas, and the
FH07 rate the coldest. However, as before, the difference between the runs with
different rate coefficients is comparable to the difference between the
realizations. The AMR simulations produce a slightly hotter core than the SPH
simulations, with a peak temperature of just over $2000~\mathrm{K}$ in the
former, compared to $\sim1700~\mathrm{K}$ in the latter. The AMR results for
ABN02 and PSS83 show a more dramatic increase in the temperature at the onset
of three-body molecular hydrogen formation than their counterpart SPH
simulations, but outside of the central molecular region (the innermost
$\sim1-10~\Msun$), both codes show good agreement.  We note also that the
radial temperature plots show better agreement to smaller scales; this is
consistent with the overall larger protostellar core in the AMR simulations.

The lower left panels of Figures~\ref{h2comp:fig:menc_multi} and
\ref{h2comp:fig:radius_multi} show the mass-weighted average molecular hydrogen
fraction, where the AMR simulations are plotted in thin lines and the SPH
simulations are plotted in thick lines.  Here, the difference between
realizations is smaller than the uncertainty introduced by the choice of
three-body rate coefficient. The physical size of the molecular cores in the
AMR and SPH simulations agree very well for the FH07 and PSS83 rates, but
disagree somewhat for the ABN02 rate, particularly for small \Hmol~fractional
abundances.  This may be a consequence of the different dissociative rate, but
that should largely be unchanged at the considered temperatures.  Additionally,
we note that the molecular core in the AMR calculations appears to be defined
by a sharper contrast as a function of radius in the ABN02 run, a result of the
lower dissociative rate in the ABN02A calculation.  The divergence between the
three rates in temperature, noted above, occurs at the radius at which the core
begins to make its transition from atomic to molecular.  The slightly higher
temperature in the ABN02A simulation at this time, $\sim2000~\mathrm{K}$, may
account for its molecular fraction approaching but not reaching
$f_{\mathrm{H}_2} = 1.0$.  All three rates, in both codes, show agreement with
the ordering of the rate coefficients themselves; the FH07 simulations have the
largest molecular core (and thus a lower density threshold for conversion,
$\sim10^{-14}~\gcc$) and the ABN02 simulations have the smallest molecular core
(and thus a higher density threshold for conversion, $\sim10^{-11}~\gcc$).

The lower right panels of Figures~\ref{h2comp:fig:menc_multi} and
\ref{h2comp:fig:radius_multi} show the mass-weighted average radial velocity
where the AMR simulations are in plotted in thin lines and the SPH simulations
are plotted in thick lines.  In the SPH simulations, the uncertainty in the
three-body rate coefficient introduces an uncertainty of roughly  $0.5 \: {\rm
km} \: {\rm s^{-1}}$ into the infall velocity. However, we note that there is
not a systematic ordering of infall velocity with rate coefficient: at an
enclosed mass of roughly $0.5 \: \Msun$, simulation ABN02S has the fastest
infall velocity of the three SPH simulations, but at an enclosed mass of $5 \:
\Msun$, it has the slowest.  The AMR simulations display a much more striking
dependence on the three-body rate coefficient. Simulations ABN02A and PSS83A
have infall velocities differing by up to $1 \: {\rm km} \: {\rm s^{-1}}$, but
broadly agree on the shape of the velocity curve, and on the location of the
peak infall velocity. Simulation FH07A, on the other hand, produces
systematically smaller infall velocities, with a peak value that is barely half
of that in the other two runs, and which occurs much further out from the
densest zone, at $r \sim 50 \: {\rm AU}$, compared with $r \sim 10 \: {\rm AU}$
in the other two runs.  Comparing the AMR and the SPH results, we see some
clear differences, as well. The magnitudes of the infall velocities in the SPH
simulations agree quite well with what is found in run FH07A, but are
significantly smaller than the velocity in runs ABN02A or PSS83A. However, the
shape of the infall velocity curves in the SPH runs agrees well with these two
AMR runs, and not so well with run FH07A. 

In Figure~\ref{h2comp:fig:menc_multi_vel} we have plotted the mass-weighted
average tangential velocity $v_{\rm tan}$ (upper panel) and the tangential velocity divided by
the Keplerian velocity (lower panel) where the Keplerian velocity is defined as
$ v_{\mathrm{kep}} \equiv \sqrt{GM_\mathrm{enc} / r} $.  As before, the AMR
simulations are in thin lines and the SPH simulations are in thick lines.  In
both sets of simulations, the uncertainty in the three-body rate coefficient
introduces significant variance in the tangential velocity between simulations,
particularly for enclosed masses greater than $\sim 1 \: \Msun$. This variance,
together with the differences in the degree of compactness of the dense
molecular core that we have already discussed, leads to an uncertainty in the
degree of rotational support of the gas, quantified by the ratio of the
tangential to the Keplerian velocities.  In the SPH simulations, which all show
a high degree of rotational support at $M_{\rm enc} > 0.5 \: \Msun$, the
variation in $v_{\rm tan}$ is relatively small, of the order of 10\%. In the
AMR simulations, which typically show less rotational support, the uncertainty
is significantly larger: between $M_{\rm enc} = 5 \: \Msun$ and  $M_{\rm enc} =
10 \: \Msun$, the ratio in run ABN02A  differs from that in run FH07A by almost
a factor of two.  We also note that the relationship between the choice of rate
coefficient and the resulting $v_{\rm tan}$ and $v_{\rm tan} / v_{\rm kep}$ is
not straightforward. At some radii, the slower rate coefficients yield larger
values, but at other radii this is reversed.  On average, the tangential
velocity in run ABN02S is higher than in runs PSS83S or FH07S, but in the AMR
runs, the situation is reversed, with ABN02A having the lowest tangential
velocity on average, and FH07A the highest.

It is also interesting to compare the results from the AMR and SPH simulations
directly. The SPH simulations clearly display a higher degree of rotational
support, consistent with the higher ratio of rotational to gravitational energy
present at the start of the simulation, compared to the ratio present in the
AMR calculation at a similar point. This also provides a simple explanation for the differences 
previously noted in the infall velocities and the central temperatures. A
higher degree of rotational support for the same enclosed mass necessarily
implies a lower infall velocity, just as we find in our simulations. Additionally,
a lower infall velocity implies a lower rate of compressional heating for the
gas at the center of the collapsing core, and hence a lower central temperature.

\subsection{Morphology}\label{h2comp:subsec:morph}

\begin{figure*}
\begin{centering}
\includegraphics[width=0.64\textwidth]{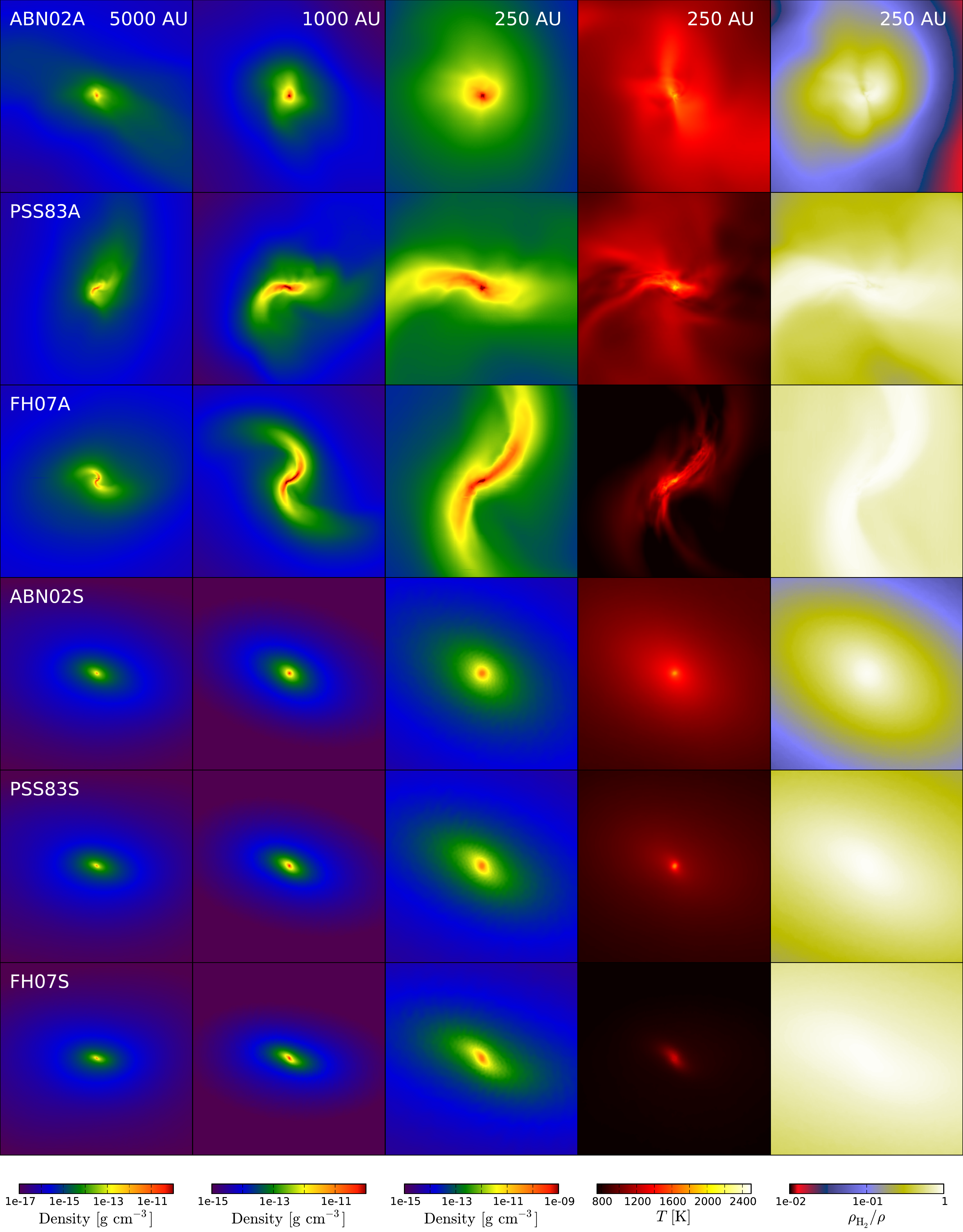}
\caption{Mass-weighted average quantities, calculated along rays at every pixel.
Rays are cast parallel to the angular momentum vector of the region in question
for every column for the AMR simulations, and parallel to the Z axis for the
SPH simulations.  Rows correspond to simulations ABN02A, PSS83A, FH07A, ABN02S,
PS83S and FH07S, and displayed values are calculated dividing mass-weighted
accumulated values by the total column density at each ray.  Column 1 shows
density with a field of view and depth of ray integration of 5000 AU, column 2
shows density with a field of view and depth of integration of 1000 AU, and
columns 3-5 show density, temperature and \Hmol fraction with fields of view and
depths of integration of 250 AU.  These images were made from snapshots of the
simulations at the time when their maximum density was $10^{-8}~\gcc$.  Color
scaling is set for each column, at ($10^{-17}~\gcc$, $10^{-10}~\gcc$) for column
1, ($10^{-15}~\gcc$, $10^{-10}~\gcc$) for column 2, ($10^{-15}~\gcc$,
$10^{-9}~\gcc$) for column 3, ($500~\mathrm{K}$, $3000~\mathrm{K}$) for column 4,
and ($10^{-2}$, $0.76$) for column 5.}
\label{h2comp:fig:fmp_z}
\end{centering}
\end{figure*}

Figure~\ref{h2comp:fig:fmp_z} shows a comparison of the morphology of the AMR
results at the final data output (when the maximum density in each simulation
is $10^{-8}~\gcc$) with those of the SPH simulations.  The morphology of the
simulations reflects the variance in the mass enclosed in the molecular cores
of the different simulations, as well as the variation in their infall
velocities.

Simulation ABN02A is mostly spherical, with little extended structure or
gaseous filaments.  The inner core of ABN02A shows no indication of angular
momentum transport through a disk-like structure at this stage in the collapse.
ABN02A also shows a considerably more extended, spherical high-temperature
region compared to the other AMR simulations. Most interestingly, the
relatively lower molecular hydrogen formation rate is evident in the smaller
molecular cloud, as the fully molecular region does not even extend for
$100~\mathrm{AU}$ from the central point of the cloud.  The temperature here is
strongly correlated with density and monotonically decreases with radius
extending outward from the center of the cloud.

However, while ABN02A is largely spheroidal, PSS83A and FH07A both show
pronounced bar-like structures.  PSS83A is spheroidal at $1500~\mathrm{AU}$,
but within $\sim100~\mathrm{AU}$ there is a bar-like structure.  The most dense
zone of the calculation is not located at the center of the bar, and the most
developed structure is at densities of $\geq5\times10^{-12}~\gcc$.  In contrast
to ABN02A, the temperature structure shows more variation at a fixed density;
specifically, in the upper left and lower right portions of the temperature
panel, we see variations at a given density.  However, while the temperature may
not track the density extremely well, the molecular hydrogen fraction appears
closely correlated with the density, just as we would expect given the steep 
$n^{3}$ density dependence of the three-body \Hmol~formation process.
Additionally, the temperature structure of the cloud encodes the shocked state 
of the gas, and small ripples are visible in the temperature structure, indicating 
both memory of the non-equilibrium molecular hydrogen formation and the 
kinematic state of the gas.  The molecular cloud also closely tracks the bar, 
although we note that this is only true for molecular hydrogen fractions of
 $\geq0.3$, as there is a substantial partially-molecular cloud outside the bar 
 structure as well.

FH07A shows the most pronounced bar, with strong density contrasts between the arms
and the embedding medium.  The bar, like that in the PSS83A simulation, is most
evident starting at densities of $\geq5\times10^{-12}~\gcc$, which in the case
of this simulation extend out to $\sim1000~\mathrm{AU}$, although they are
compressed along the axis of the bar.  This structure is also evident in the
temperature structure, but most importantly we note that it is embedded within
the molecular cloud, which extends well beyond the highest density regions.

We see no evidence of advanced stages of fragmentation in any of the AMR
simulations; however, the extremely compressed spiral arms of FH07A may become
gravitationally unstable at later times, depending on the subsequent evolution of
the angular momentum. Simulation PSS83A shows a broadly spherical mass-distribution 
on the scale of $2500~\mathrm{AU}$, but a substantially less symmetrical, nearly 
cardiod-shaped mass-distribution on scales of $\sim500~\mathrm{AU}$.

The SPH simulations show considerably less of the fine detail apparent in the
AMR simulations, due primarily to their considerably lower effective resolution
on these scales. All three SPH runs show a largely spheroidal morphology, but
the same basic trend is visible here as in the AMR simulations. Run ABN02S is
the most spherical, while runs PSS83S and FH07S display increasingly flattened
gas distributions, and the beginning of a bar is visible in the FH07S run.  Run
ABN02S has the highest internal temperature, as reflected in the radial
profiles, as well as the most spherical shape in the inner $250~\mathrm{AU}$
cloud; however, we note that it is oblate at larger scales
($\sim1000~\mathrm{AU}$), and that the molecular hydrogen morphology tracks the
density structure.  As expected from the radial profiles discussed above, and
in keeping with the results from ABN02A, the ABN02S molecular hydrogen cloud is
the smallest of the clouds of the three SPH simulations.  

\subsection{Accretion Rates}\label{h2comp:subsec:acc}

\begin{figure*}
\plotone{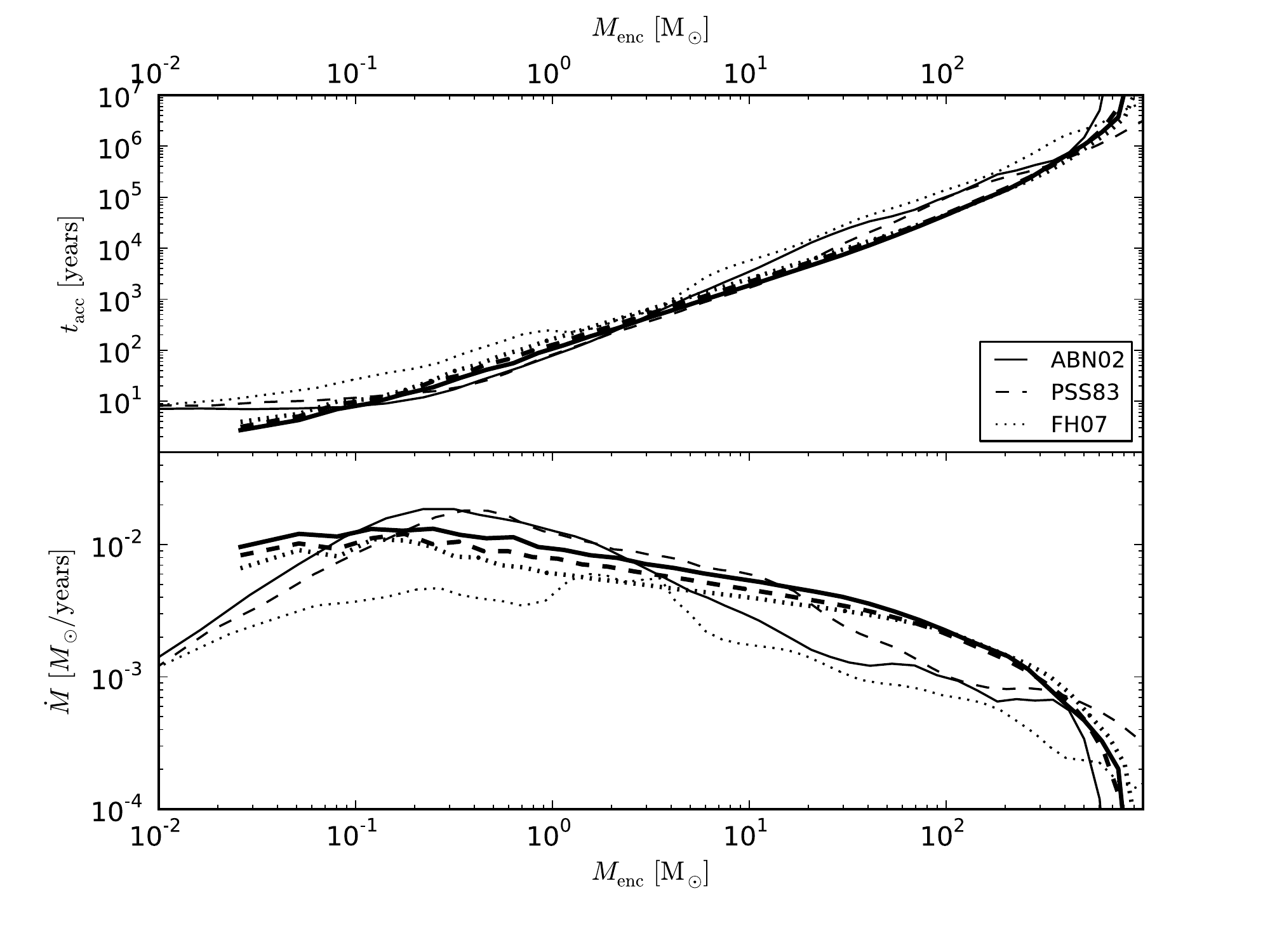}
\caption{Radially-binned, spherically averaged accretion times (top) as a
function of enclosed mass, and the accretion rates (bottom) for each mass
shell.  These values were calculated at the final output time of the simulation.
Thin lines correspond to the AMR simulations and thick lines correspond to the
SPH simulations.}
\label{h2comp:fig:menc_multi_acc}
\end{figure*}

In Figure~\ref{h2comp:fig:menc_multi_acc}, we plot the accretion rates and accretion
times as a function of enclosed mass, calculated at the final output time of both
sets of simulations.  

In the top panel, we plot the accretion time.  Taking $v_r$ as the radial
velocity at the radius $r$ and $M_{\mathrm{enc}}$ as the mass enclosed within
a given radius, the accretion time is given by $$ t_{\mathrm{acc}} =
\frac{M_{\mathrm{enc}}}{4\pi\rho v_r r^2}.  $$ For enclosed masses of less than 
$\sim2~\Msun$, the two sets of simulations show good agreement in the ordering of 
the accretion times; the FH07 runs have the longest accretion times, then the 
PSS83 runs, with the ABN02 runs having the shortest accretion times.  However,
for enclosed masses greater than $2~\Msun$,  the AMR simulation ABN02A increases 
with respect to the mean, and the ordering between the SPH and AMR simulations 
is no longer identical.  Additionally, at the specific mass of $\sim2~\Msun$ the six 
simulations converge on a time scale of $\sim300~\mathrm{years}$ for that mass 
to accrete onto the most dense zone, assuming direct infall.  We note that this is 
approximately the mass scale of the molecular cloud in all six simulations, and 
thus convergence between the rates at this point is not surprising.

Considering only the accretion times between enclosed masses of $10-100~\Msun$,
we see variation of up to a factor of two within the AMR simulations, but
relatively close convergence within the SPH calculations.  This is not
surprising: as discussed earlier, the AMR calculations in general show much
greater variation of morphology and radial velocity as a function of the
three-body rate used.  However, at enclosed masses of $\sim100~\Msun$~we note
that the AMR simulations show good agreement, diverging slightly at greater
enclosed mass, which we can attribute to a change in the settling of the cloud
as a result of time delay between the simulations.

The accretion rates, like the accretion times, show substantial variation at
all mass scales.  In particular, the ABN02A simulation has the highest
accretion rate, peaking at $0.02~\Msun~\mathrm{yr}^{-1}$ at
$0.2~\Msun$~enclosed.  Simulations ABN02A and PSS83A show steeper curves in the
change of accretion rate with enclosed mass than the SPH calculations, but
FH07A is generally more irregular as a function of enclosed mass; we attribute
this to its highly aspherical collapse, in contrast to the more spheroidal
ABN02A and PSS83A calculations.

%%% Discussion %%%

\section{Discussion}\label{sec:discussion}
Our simulations have shown that as we decrease the three-body \Hmol~formation
rate, we find several systematic changes in the properties of the collapsing
gas. The molecular region at the center of the collapse becomes significantly
smaller, and the density structure of this central region changes, becoming
less compact and more spherical, with less small-scale structure. In addition,
the time taken for the gas to collapse becomes longer. These general features
appear to be independent of our choice of simulation technique or dark matter
halo, although our two sets of simulations do show some disagreements on the
magnitude of these differences. 

These general differences are simple to relate to the microphysics of the gas.
It should be no surprise that if we reduce the rate of \Hmol~formation, then we
find that the molecular fraction at a given density decreases. It is also not
particularly surprising that if we reduce the amount of \Hmol~present in gas of
a given density, thereby reducing the ability of the gas to cool efficiently,
then we find that the gas becomes warmer than it would be if more \Hmol~were
present (with attendant effects, e.g.\ \citealt{Turk2010a}.) What is
potentially more surprising is that the large changes in \Hmol~abundance that
we see between radii of a few hundred and a few thousand AU do not have a
greater effect on the temperature structure of the gas.  However, this is in
part a consequence of the steep dependence of the \Hmol~cooling rate on
temperature: an increase in temperature of only a few hundred Kelvin can offset
a considerable reduction in the \Hmol~abundance. In addition, chemical heating
of the gas is also significantly lower in simulations with smaller three-body
rates, for obvious reasons.

The consequences of the less efficient cooling and the delayed collapse of the
gas are less straightforward.  Naively, we might expect that if the collapse is
delayed, then there will be more time for the outward transport of angular
momentum, resulting in a gas distribution with less rotational support, and
hence a higher infall velocity. In practice, although some of our runs support
this picture (e.g.\ run FH07A), others suggest that this is an
oversimplification. Reducing the three-body \Hmol~formation rate, and hence
delaying the collapse, leads to less rotational support at some radii, but it
also leads to greater support at other radii. Moreover, the details appear to
depend on the particular realization of collapse studied. Nevertheless, the
general lesson that we can learn from the simulations is that the uncertainty
in the three-body \Hmol~formation rate significantly limits our ability to model
the density, temperature and velocity structure of the gas close to the center
of the collapse.

The consequences of this uncertainty are not easy to assess, given our lack of
knowledge regarding exactly what happens following the formation of the initial
Population III protostar. Many models for the later accretion phase presuppose the 
formation of an accretion disk around this protostar 
\citep[e.g.][]{TanMcKee2004,2010MNRAS.403...45S,Turk2010c,Clark2010S} Our simulation 
results clearly demonstrate that the mass assembly history of any such disk would
be uncertain, given the uncertainties in the collapse time and collapse structure of
these clouds.  At one extreme, simulation FH07A forms an extremely strong spiral-bar
structure with strong rotational support; at the other extreme, ABN02A is
almost a spherically-symmetric collapse with a markedly higher accretion time.

At very early times during the formation of the central protostar and the initial
assembly of the disk, we would not expect to see significant differences. The
thermodynamics of the very dense gas that forms the initial protostar is controlled
by the dissociation of molecular hydrogen, and is independent of our choice of
three-body \Hmol~formation rate \citep{2008Sci...321..669Y}. However, the
larger-scale regions in which we see more substantial deviations in density,
temperature and velocity structure, will control the rate at which gas is fed
onto the disk, and thence onto the newly formed hydrostatic core. Higher
accretion rates onto the disk may make it more unstable, and hence more likely
to fragment (e.g.  \citealt{2010ApJ...708.1585K}).  Higher accretion rates onto
the protostar may dramatically change the character of the radiative feedback
from it, and hence may substantially alter its final mass
\citep{2003ApJ...589..677O}.  Furthermore, the speed of larger-scale accretion
will change the mass and radius at which the initial collapse of the gas
becomes fully  adiabatic.  Entering the adiabatic phase of collapse at lower
densities leads to a larger radii for adiabatic compression and thus a larger
mass scale.

Of potentially more relevance to the final mass of Pop. III stars is the
variation in the collapse times discussed in \S \ref{sec:timing}. It has been
demonstrated that the gas in minihalos may contain sufficient structure to
allow it to fragment during the collapse \citep{2009Sci...325..601T}. If this
occurs, then rather than all of the mass accreting onto one central protostar,
it must now be shared amongst the multiple stars making up the protostellar
system. However the ability of structure to survive will depend strongly on
both the temperature of the gas and the time that it takes to collapse, since
together these determine the time during which sound waves can act to remove
the anisotropies. As discussed in \S \ref{sec:timing}, simulations FH07A and
FH07S collapse between 20,000 and 140,000 years faster than simulations ABN02A
and ABN02S.  By comparison, at densities of around $10^{-16}~\gcc$, the point
at which the three-body reactions start to become important, the free-fall time
in the gas is only $\sim 6700$ yr. We would therefore expect the details of the
fragmentation to depend on the rate at which the \Hmol~forms, as well as the
details of calculations of the optically-thick cooling rate of \Hmol.

Finally, it is also clear from our study that the differences between different
numerical realizations of Population III star formation are often as large or
larger than the uncertainties introduced by our lack of knowledge regarding the
three-body \Hmol~formation rate. This fact limits our ability to make general
statements about the impact of the rate coefficient uncertainty on e.g.~the
Population III IMF.  However, we can in principle address this by performing a
large ensemble of simulations, so as to fully explore the entire parameter
space of initial conditions (see e.g.\ \citealt{oshea07a}, or
\citealt{Turk2010d}), but we cannot eliminate the rate coefficient
uncertainties in this fashion.
 
%%
%% Conclusions
%%

\section{Conclusions}\label{sec:conclusions}
We have shown that the uncertainty in the rate coefficient for the three-body
formation of molecular hydrogen from atomic gas has a significant effect on 
the details of the collapse of primordial star-forming clouds in the high-density
regime. The differences in outcome brought about by choosing a different rate
coefficient are most dramatic in what is typically considered the inner cloud, 
where the protostellar disk will begin to form, but we can reasonably expect these 
changes to propagate outwards over the course of the accretion onto the 
protostellar core.

The density scale at which molecular hydrogen forms dramatically changes the
chemical makeup, morphology and velocity structure of the gas in the inner
regions of the protogalactic gas cloud.  In the isothermal collapse model, the
accretion rate is governed by the sound speed; therefore a higher temperature,
as a result of later molecular formation, results in a higher accretion rate.

The differences between runs with different three-body \Hmol~formation rates
are comparable to the differences between different realizations of primordial
protostellar collapse. However, the latter issue can be addressed simply by
simulating a large number of different realizations (e.g.\ \citet{Turk2010d}),
which will allow the natural variance in collapse rates, degree of rotational
support, etc.\ to be studied and quantified.  The uncertainty in the outcome of
collapse caused by our poor knowledge of the three-body \Hmol~formation rate
coefficient cannot be so easily dealt with, and represents a major limitation
on our ability to accurately simulate the formation of the first stars in the
Universe. Furthermore, recent suggestions that metal-free star-forming clouds
could fragment into multiple protostellar cores
\citep{2009Sci...325..601T,2010MNRAS.403...45S,Clark2010S} place an increased
urgency on understanding the chemistry of primordial gas.  Changes in the
structure of the molecular cloud, on the scales of a few thousand
$\mathrm{AU}$, whether as a result of time-delay in the collapse or a change in
the thermal structure, could induce or suppress fragmentation.

Finally, we would like, at this point, to be able to recommend a particular
rate coefficient as the best available choice, but the truth of the matter is
that there seem to be few compelling reasons to prefer one choice over another
from amongst the available rates.  The most conservative choice would probably
be to disregard the two most extreme choices (the rates from \citetalias{ABN02}
and \citetalias{2007MNRAS.377..705F}) and  take one of the rates that produces
intermediate values at low temperatures, such as the rates from
\citetalias{PSS83} or  \cite{2008AIPC..990...25G}. Of these intermediate rates,
we prefer the latter, as it is based on a relatively recent calculation of the
collisional dissociation rate of \Hmol, rather than on an extrapolation from
data that is more than forty years old. Nevertheless, even this choice is at
best a stopgap until more accurate values for the rate coefficient become
available.

Unfortunately, the prospects of this situation improving in the near future 
are dim. To the best of our knowledge, there are currently no experimental
groups capable of studying this process at cosmologically relevent temperatures.
Indeed, such measurements may be just beyond current experimental capabilities, 
given the required combination of high atomic number density and high temperature
(D.~W.~Savin, private communication). It seems likely that this will remain an
unavoidable uncertainty in studies of population III star formation for some 
time to come.

\acknowledgments
M.J.T.\ acknowledges support by NASA ATFP grant NNX08AH26G and NSF AST-0807312.
P.C.C.\ acknowledges support by the {\em Deutsche Forschungsgemeinschaft} (DFG)
under grant KL 1358/5. R.S.K.\ acknowledges financial support from the {\em
Landesstiftung Baden-W\"urrtemberg} via their program International
Collaboration II (grant P-LS-SPII/18) and from the German {\em
Bundesministerium f\"ur Bildung und Forschung} via the ASTRONET project STAR
FORMAT (grant 05A09VHA). R.S.K.\ furthermore acknowledges subsidies from the
DFG under grants no.\ KL1358/1, KL1358/4, KL1358/5, KL1358/10, and KL1358/11,
as well as from a Frontier grant of Heidelberg University sponsored by the
German Excellence Initiative.  R.S.K.  also thanks the Kavli Institute for
Particle Astrophysics and Cosmology at Stanford University and the Department
of Astronomy and Astrophysics at the University of California at Santa Cruz
for their warm hospitality during a sabbatical stay in spring 2010.
V.B.\ acknowledges support from NSF grant AST-0708795 and NASA ATFP grant
NNX08AL43G. Part of the simulations were carried out at the Texas Advanced
Computing Center (TACC), under TeraGrid allocation TG-AST090003.  Part of the
simulations were conducted at the SLAC National Accelerator Laboratory on the
Orange cluster.

%\bibliography{mjt}
%\IfFileExists{\jobname.bbl}{}
% {\typeout{}
%  \typeout{******************************************}
%  \typeout{** Please run "bibtex \jobname" to optain}
%  \typeout{** the bibliography and then re-run LaTeX}
%  \typeout{** twice to fix the references!}
%  \typeout{******************************************}
%  \typeout{}
% }

\end{document}